 \documentclass[10pt,preprint]{aastex}  
 \usepackage{natbib}
\def\gapprox{\lower.4ex\hbox{$\;\buildrel >\over{\scriptstyle\sim}\;$}}
\def\lapprox{\lower.4ex\hbox{$\;\buildrel <\over{\scriptstyle\sim}\;$}}

\shortauthors{ASCHWANDEN}
\shorttitle{Exoplanet Predictions}

\begin{document}

\title{         Exoplanet Predictions Based on Harmonic Orbit Resonances } 

\author{        Markus J. Aschwanden$^1$}

\affil{		$^1)$ Lockheed Martin, 
		Solar and Astrophysics Laboratory, 
                Org. A021S, Bldg.~252, 3251 Hanover St.,
                Palo Alto, CA 94304, USA;
                e-mail: aschwanden@lmsal.com }

\and	

\author{	Felix Scholkmann$^2$ 	}

\affil{		$^2)$ Research Office for Complex Physical and Biological Systems,
		Mutschellenstr. 179, 8038 Z\"urich, Switzerland;
		e-mail: felix.scholkmann@gmail.com		}		

\begin{abstract}
The current exoplanet database includes 5454 confirmed planets 
and candidate planets observed with the KEPLER mission. We find
932 planet pairs from which we extract distance and orbital period
ratios. While earlier studies used the Titius-Bode law or a
generalized version with logarithmic spacing, which both lack 
a physical model, we employ here the theory of harmonic orbit
resonances, which contains quantized ratios instead, to explain the 
observed planet distance ratios and to predict undetected exoplanets. 
We find that the most prevailing
harmonic ratios are (2:1), (3:2), and (5:3), in 73\% of the cases,
while alternative harmonic ratios of (5:4), (4:3), (5:2), (3:1) 
occur in 27\% of the other cases. Our orbital predictions includes 
171 exoplanets, 2 Jupiter moons, one Saturn moon, 3 Uranus moons,
and 4 Neptune moons. The accuracy of the predicted planet distances 
amounts to a few percent, which fits the data significantly better
than the Titius-Bode law or a logarithmic spacing. This information 
may be useful for targeted exoplanet searches with Kepler data
and to estimate the number of live-carrying planets in habitable zones.
\end{abstract}

\section{	INTRODUCTION			}

The recent discoveries of exoplanets has currently amassed a 
catalog of over 5000 exoplanets (Han et al.~2014), which contains 
the names of the host stars, the semi-major axes, and orbital periods 
of the associated exoplanets (see website: {\sl exoplanets.org}). 
The unprecedented statistics of these orbital parameters allows us
for the first time to determine whether the planets are distributed
in random distances from the central star, or in some systematic order,
as it is expected from physical models of harmonic resonance orbits
(Peale 1976). Data analysis of statistical exoplanet databases
therefore provides key information on the physical process of
the formation of planetary systems, as well as on the number and 
statistical probability of planets that are located in habitable 
zones around stars.

Orbital resonances in the solar system have been known for a long time
(e.g., see review by Peale 1976). A particular clean example is given
by the three Galilean satellites of Jupiter (Io, Europa, Ganymede),
for which Laplace (1829) demonstrated their long-term stability 
based on celestial mechanics perturbation theory. Other examples
are the gaps in the rings of Saturn, the resonances of asteroids
with Jupiter (Trojans, Thule, Hilda, Griqua, Kirkwood gaps), as 
well as the harmonic orbital periods of the planets in our solar
system. Low harmonic (integer) numbers of orbital periods
warrant frequent conjunction times, which translates into frequent
gravitational interactions that can stabilize a three-body system
of planets (or asteroids). The stability of orbital resonances
has been explained in terms of the libration of coupled pendulum
systems (Brown and Shook 1933). Physical descriptions of 
celestial resonance phenomenona are reviewed in Peale (1976). 

Empirical models of planet distances, such as the Titius-Bode law,
or the generalized Titius-Bode law, are not consistent with harmonic
orbit ratios, and lack a physical model. It is therefore
imperative not to use such empirical laws for the prediction of 
exoplanets, but rather physical models based on harmonic orbital resonances,
which we pursue in this study. Previous predictions of (missing)
exoplanets are based on the original Titius-Bode law (Cuntz 2012; 
Poveda and Lara 2008; Qian et al.~2011), the logarithmic spacing 
of the generalized Titius-Bode law (Bovaird and Lineweaver 2013; 
Bovaird et al.~2015; Huang and Bakos 2014), an exponential 
function (Lovis et al. 2011), or a multi-modal probability distribution
function (Scholkmann 2013). A recent study analyzed the statistical
distribution of mean motion resonances and near-resonances in
1176 exoplanet systems, finding a preference for (3:2) and (2:1)
resonances in the overall, but large variations of the harmonic ratio
for different planet types, grouped by their semi-major axis size,
their mass, or host star type (Ghilea 2015). 

In this study
we identify a complete set of principal harmonic numbers that explain
all observed exoplanet distances and infer the statistical probability
of the harmonic number ratios found in exoplanets. Based on this 
analysis we present predictions of so far un-detected exoplanets
in 190 exoplanet systems. 
We describe the data analysis method and the observational results in 
Section 2, followed by a discussion conclusions in Section 3.

\section{	OBSERVATIONS AND DATA ANALYSIS METHOD		}

\subsection{	Data Set Statistics				}

We access the web-based exoplanet database {\sl http://exoplanets.org} and 
find at the time of this analysis (2017 May 10) a total of 5454 
detected exoplanets,
which includes confirmed planets as well as candidate planets from the
{\sl Kepler} mission. Most of the exoplanets represent single
detections in a star system, which is the case in 3860 cases.
Double planet detections occur in 472 star systems, triple-planet
systems in 128, quadruple-planet systems in 44,
quintuple-planet sytems in 14, and sextuple-planet systems
in 2 star systems. The largest planet systems have 7 detections of planets,
such as the KIC-11442793 and the TRAPPIST-1 system (Gillon et al.~2017).
We updated the TRAPPIST-1 system, with 3 stars given in the exoplanet
data base, to 7 stars given in Gillon et al.~(2017), analyzed also by
Pletser and Basano (2017) and Scholkmann (2017).
All these exoplanet detections are associated with a total of 4522
different stars (Table 1).

For our analysis we are most interested in studying harmonic resonance 
orbit ratios, which requires at least 2 neighbored exoplanets per star,
which amounts to a total of 622 exoplanets, or a number of 932 planet pairs.
We attempt also predictions of missing planets, which requires at least
3 planets per star, which is feasible in 190 exoplanet systems.

\subsection{	Orbital Period Ratio Distribution 		}

An important result of our analysis is the distribution of orbital period
ratios $Q_i=P_{i+1}/P_i$, where $P_i$ and $P_{i+1}, i=1,...,n_p-1$ are 
the orbital time periods in a planetary system with $n_p$ planets. 
These orbital periods $P_i$
are directly related to the semi-major axis of the planets (i.e., their
mean distance $D$ to the center of the host star), according to
Kepler's 3$^{rd}$ law,
\begin{equation}
	D = P^{2/3} \ .
\end{equation}
We plot a histogram of orbital period ratios $Q_i$ for 
all $N=932$ planet pairs in Fig.~1 (top panel), sampled in a range of 
$P=[0.5,3.5]$ with a histogram bin width of $dP=0.04$. These period ratios are
defined from the ratios of the outer planet distance $P_{i+1}$ to 
the inner planet distance $P_i$, which is always larger than unity 
by definition, i.e., $P_i > 1$, since the inner planets rotate faster around
the Sun than the outer planets. We are adding error bars $\sigma_N(P_i)
\propto \sqrt{N_i}$ in Fig.~1 (top panel) as expected from Poisson statistics. 

What is striking about the period distribution $N(P)$ shown in Fig.~1 
(top panel) is that the periods do not follow a smooth (approximately
log-normal) distribution, but rather
exhibit significant peaks at the harmonic values of $Q=(3:2)=1.5$
and at $P=(2:1)=2.0$, which involve the lowest possible harmonic 
numbers 1, 2, and 3. In a previous study on our own solar system and the moons
around planets we noticed a dominance of five low harmonic ratios,
i.e., (3:2), (5:3), (2:1), (5:2), and (3:1) (Aschwanden and McFadden 2017).
Interestingly, we find now that exoplanets confirm the preponderance
of harmonic ratios (3:2) and (2:1), while the other 
ratios (5:3), (3:1), and (5:2) are less significant.

The distribution of harmonic ratios drops steeply from $Q=1.5$
towards a value of $Q > 1.0$ at the low side, while the upper end 
extends all the way to $Q \approx 1229$. The extended tail at the 
upper end can be explained by a large number of missing (undetected) 
planets, where the most extreme ratio applies to the case where the
innermost and the outermost planet is detected only. For our solar
system, this maximum ratio would be the orbital period ratio
between Mercury and Pluto, i.e., $Q_{max}=248$ yrs / 0.24 yrs 
$\approx 1033$.

\subsection{	Orbital Period Ratios of Large Planet Systems	}

In Fig.~2 we show the orbital distances $D$ for the 18 largest 
exoplanet systems (in alphabetical order), which includes all 
cases with 5, 6, or 7 planets per star. Each panel in Fig.~2
displays the predicted planet distance (from the central star)
to the observed planet distance. While we quote
harmonic ratios of the orbital periods throughout the paper,
the corresponding distances (as shown in Fig.~2) are simply
calculated from Kepler's $3^{rd}$ law (Eq.~1). 
If we limit ourselves to a maximum orbit ratio of $Q \le 3.0$, we
find that all observed orbital period ratios in this subset
(shown in Fig.~2) can be explained by 7 small harmonic ratios 
in the range of
(5:4)=1.25, (4:3)=1.333, (3:2)=1.5, (5:3)=1.667, (2:1)=2.0,
(5:2)=2.5, (3:1)=3.0, which we define as ``principal harmonic ratios''.
From these largest 18 systems shown in Fig.~2
we find that half of them are ``gap-free exoplanet systems'',
which means that all detected period ratios $Q_i$ can be explained with
these principal harmonic ratios in the range of
$1.25 \le Q \le 3.0$, such as Kepler-11, Kepler-154, Kepler-292,
Kepler-33, Kepler-444, Kepler-55, Kepler-80,
Kepler-84, and TRAPPIST-1. In all other cases we find some 
``gaps'', which are manifested by orbital period ratios of $Q > 3.0$.
These gaps are likely to be produced by missing (undetetected) planets 
(marked with red diamonds in Fig.~2, such as 55-Cnc, HD-10180, 
KIC-11442793, Kepler-169, Kepler-186, Kepler-20, 
Kepler-32, and Kepler-62. Since the largest exoplanet systems
have 7 members so far, while our own solar system hosts 10 planets,
it is likely that there are more missing exoplanets outside the 
so far detected gap-free sequences.

\subsection{	Strategy for Predicting Missing Exoplanets 		}

Previous studies developed strategies to predict missing (undetected)
exoplanets from the generalized Titius-Bode law, which assumes
a constant geometric progression factor $Q$, i.e., $T_{i+1}/T_i = Q$,
(e.g., Bovaird and Lineweaver 2013). In a recent study, however,
we found that the observational data support quantized harmonic 
resonance ratios, such as $Q=1.5, 1.667, 2.0, 2.5, 3.0$ 
(Aschwanden and McFadden 2017), rather than a constant progression factor $Q$.
Such a quantization thus has to be included in a search strategy
of missing (undetected) exoplanets. On the other side, the
generalized Titius-Bode law is simpler to fit to data, because
it is controlled by a single free parameter $Q$, while fitting
a harmonic resonance model is more ambiguous due to a larger
number of quantized parameters $Q_i$. 

The problem can be formulated as an optimization problem of predicting 
a gap-free sequence of harmonic ratios,
\begin{equation}
	P_{i+1}^{pred} = P_i^{pred} \times Q_i, \qquad i=1,..., n_p-1 \ ,
\end{equation}
where the $n_p$ orbital period ratios $Q_i$ are drawn from the
range of principal harmonic ratios in such a way that the
observed subset of $n_{det} \le n_p$ orbital periods $P_i^{obs}$ of the 
detected exoplanets matches the predicted periods $P_i^{pred}$.

In the following we develop a search strategy that is consistent
with harmonic resonance orbits and consists of the following steps:
(1) We define a range of 7 {\sl principal low harmonics}: 
(5:4)=1.25, (4:3)=1.333, (3:2)=1.5, (5:3)=1.667, (2:1)=2.0,
(5:2)=2.5, (3:1)=3.0. Neighbored exoplanet orbit pairs with 
ratios in the range of $1.1 \lapprox Q_i \lapprox 3.1$ are attributed 
to the closest principal harmonic ratio. 
(2) Orbital ratios in the range of $1.0 \le Q_i \lapprox 1.1$
are interpreted as planet pairs in identical harmonic resonance
zones, which are found only for very small bodies (like asteroids
or moons of Jupiter and Saturn with diameters of $d < 500$ km);
(3) Detected planet pairs with orbital periods of
$Q \gapprox 3.1$ are interpreted as gaps with one or more missing 
(undetected) planets. Since a gap $Q_{obs}$ has to 
be filled by $n_{miss}$ planets, i.e., $Q_{obs} \approx (Q_{gap})^{(n_{miss})}$, 
we determine the number $n_{miss}$ of missing planets from the 
observed interval $Q_{obs}=T_{i+1}/T_i$ by 
\begin{equation}
	n_{miss} = Round \left[ {\log{Q_{obs}} \over \log{Q_{gap}}} 
		   \right] \ ,
\end{equation}
where {\sl Round} represents a rounding function to the next integer number.
The characteristic ratio $Q_{gap}$ is estimated from the 
best-fitting value of the 5 most frequent principal harmonic ratios, 
i.e., $Q_{gap}$=1.5, 1.667, 2.0, 2.5, and 3.0. The best-fitting value
$Q_{gap}$ is simply determined by minimizing the mean deviation $\sigma_P$ 
between the observed ($P_i^{obs}$) and the predicted orbital periods ($P_i^{pred}$),
\begin{equation}
	\sigma_P = {1 \over n_{obs}} \sum_{i=0}^{n_{obs}} 
	{|P_i^{pred}-P_i^{obs}| \over P_i^{obs} } \ ,
\end{equation}
which translates into a mean deviation $\sigma_D$ of planet distances $D$
according to Kepler's law (Eq.~1), 
\begin{equation}
	\sigma_D = {1 \over n_{obs}} \sum_{i=0}^{n_{obs}} 
		{|D_i^{pred}-D_i^{obs}| \over D_i^{obs} } \ ,
	         = \sigma_P^{2/3} \ .
\end{equation}
An additional constraint is the estimated maximum number of planets per system,
which we choose to be $n_{p,max}=11$, based on the number of planets detected
in our solar system ($n_p=10$ including the asteroids), and the number
of Jupiter moons ($n_p=9$) or Saturn moons ($n_p=10$) that fit the scheme
(see Appendix A for our solar system and planetary moon systems). 

These best-fit mean deviations $D_{pred}/D_{obs}-1=\sigma_D$ are listed in
the 18 panels in Fig.~2. We see that these best-fit solutions of harmonic
ratios match the observed orbital period ratios typically within 
accuracy of $\approx 1\%-5\%$. The most accurate harmonic
ratios with an accuracy of $\approx 1\%$ are found for the planetary systems of 
Kepler-169, Kepler-32, Kepler-444, and TRAPPIST-1.

Of course, the estimated
missing planets bear some ambiguity because the ratios in
subsequent intervals in a gap can be permutated arbitrarily. 
For instance, if we take out a planet between two resonance 
zones of $Q_{i+1}=(3:2)$ and $Q_i=(2:0)$, the permutated 
sequence $Q_{i+1}=(2:0)$ and $Q_i=(3:2)$ yields an identical 
solution, and thus the predicted resonance 
at the location of the filled gap is only accurate to 
$(1.5+2.0)/2 = 1.75 \pm 0.25$, which is attributed to the closest 
admissible resonance ratio of $(5:3)=1.667$ and yields a 
combined period ratio of $(5:3)\times (5:3) = 2.778$ in the gap, 
which is within 8\% of the correct gap value $(3:2)\times (2:1)=3.0$. 

\subsection{	Statistical Probability of Quantized Periods 	}

After we corrected each detected exoplanet sequence of orbital
periods $Q_i, i=1,...,n_p-1$ per exoplanet system or star (for
exoplanet systems with more than 3 detected exoplanets), by
predicting missing exoplanet detections according to the
algorithm described above, we can sample the distribution of
predicted orbital periods, which is shown in Fig.~1 (bottom panel).
The predicted distribution of orbital period ratios
is now, as a consequence of the applied prediction scheme,
quantized according to the 7 principal harmonic ratios, while
no ratio is found outside the range of $1.25 \le Q_i \le 3.0$.
What this quantized distribution reveals moreover, is the
statistical probability for the individual harmonic ratios.
According to the result shown in Fig.~1 (bottom panel),
the highest probability is for the harmonic ratios of
(3:2), (5:3), and (2:1), while the other
four harmonic ratios are significantly rarer. The relative
probabilities for each of the 7 principal harmonic ratios are
(in order of decreasing probability):
26\% for a ratio of (3:2),
25\% for a ratio of (2:1),
22\% for a ratio of (5:3),
14\% for a ratio of (5:2),
6\% for a ratio of (3:1), 
4\% for a ratio of (4:3), and
3\% for a ratio of (5:4).  

\subsection{	Catalog of Exoplanet Predictions	}

In Fig.~3 we list the planet distances of all 190 exoplanet 
systems with more than 3 planets. The observed period ratios amount
of the 190 exoplanet systems amount to a total of
$N_{det}=128 \times 3 + 44 \times 4 + 14 \times 5 + 2 \times 6 + 2 \times 6
=654$ values (see statistics in Table 1). 
The number of predicted exoplanets for this sample are
marked with red numbers in Fig.~3, amounting to a total of 171 predictions.
These predicted distances of undetected exoplanets can be
used in targeted searches with Kepler data or others.

\section{	DISCUSSION AND CONCLUSIONS		}

The knowledge of planet distances from their host star provides us
several important pieces of information. One central insight that
can only be proven with reliable planet distance ratios is the 
physical formation process of planetary systems. A leading theory
in this respect is the harmonic orbit resonance concept, which 
implies that multiple planets entrain into long-term stable orbits
only once their orbits achieve a harmonic ratio with low numbers.
Our analysis of 932 exoplanet pairs yielded evidence that the
prevailing harmonic ratios are (2:1), (3:2), and (5:3), which 
dominate in 73\% of all exoplanet pairs, consistent with the 
finding of Ghilea (2015), while less additional harmonic ratios 
of (5:4), (4:3), (5:2), and (3:1) occur also, but are less frequent 
(in 27\% of all cases). This relative probability of harmonic ratios
in planet orbits is a quantitative result that can be tested with 
numerical simulations of n-body gravitational perturbation theory.

What is the predictive power of this exercise for missing exoplanets?
Previous studies employed the generalized Titius-Bode law (e.g.,
Bouvaird and Lineweaver 2013), which assumes a constant geometric
progression factor (or orbit period progression factor $Q$ by 
applying Kepler's 3${rd}$ law), also called logarithmic spacing.
A constant progression factor has mostly been used for sake of 
mathematical simplicity, because it has
only one free parameter, but it has no physical justification, and 
is actually inconsistent with harmonic orbital resonance theories, 
which predict quantized values of low harmonic number ratios, 
such as our set of 7 principal harmonic numbers. Therefore, 
the prediction of missing planets requires an algorithm that
is based on low harmonic number ratios, which is more difficult
to fit in the presence of gaps than a constant progression factor 
for logarithmic spacing.
Consequently, in fitting a complete planet system without gaps
(of missing undetected planets), each planet orbit period ratio
has up to 7 free parameters (corresponding to the principal
low harmonic ratios). Although this creates some ambiguity, since
the best solution is insensitive to permutations of interpolated
orbit period ratios, we eliminated this ambiguity by minimizing
the difference between observed and modeled planet distances.
In summary, the prediction of missing planets is based on solutions 
of orbital period sequences that contain only harmonic number ratios,
rather than a constant logarithmic spacing. 

The orbital predictions obtained in this study includes 2 Jupiter
moons, one Saturn moon, 3 Uranus moons, 4 Neptune moons, and
171 exoplanets (Fig.~2 and Table in Fig.~3). The accuracy of the predicted
planet distances amounts to a few percent. A previous study
has shown that the harmonic orbit resonance model with quantized 
values fits the solar system and lunar systems better
than the assumption of uniform logarithmic spacing
(Aschwanden and McFadden 2017). This information may be
useful for targeted searches of exoplanets with Kepler data,
since it reduces the search parameter space. Moreover, it allows
us to estimate the number of habitable planets in each exoplanet
system (Chandler et al.~2016). 

The existence of quantized values in planetary distances represents
a system with (non-random) order, which falls into the category
of self-organizing systems (Aschwanden and McFadden 2017).
Self-organizing systems are characterized by regular geometric 
patterns that result from frequent local interactions in an initially
disordered system (e.g., the libration of coupled pendulums). 
The principle of self-organization, however,
should not be confused with the concept of {\sl self-organized
criticality} (Bak et al. 1987) in nonlinear dissipative systems, 
which is common in astrophysics also (Aschwanden et al.~2016). 

\section*{	APPENDIX A: Solar System and Planetary Moon Systems	}

As a consistency test we apply our prediction algorithm for missing
exoplanets (Section 2.4) to our solar system and the moon systems
described in a previous study (Aschwanden and McFadden 2017).
The results are shown in Fig.~2c. 

First we use the planet distances from Sun center for all 10 planets
of our solare system, including the asteroid Ceres, shown in the
panel labeled with Sun-10. Since no harmonic ratio larger than (3:1)
occurs, our algorithm predicts no missing planet. In the second panel
(labeled with Sun-9) we eliminate the asteroid Ceres, which is then
detected as missing planet with our algorithm, because the harmonic
ratio between Mars and Jupiter exhibits a ratio of $\approx 4.0$ that
is significantly larger than the largest admissible principal harmonic
ratio of 3.0. In the panel Sun-8 we eliminate Pluto in addition, which
mimics the situation before 1930. In the panel Sun-7 we eliminate
Neptune (before 1846), and in panel Sun-6 we eliminate Uranus (before
1781). In all four cases Sun-9, Sun-8, Sun-7, and Sun-6, our algorithm
consistently flags only the asteroid Ceres, because in each 
situation the outermost planet is removed, while the remaining planets
represents a gap-free sequence. 

In panel Jupiter-7 (in Fig.~2c) we sample all Jupiter moons with a
diameter of $d > 100$ km, namely Amalthea, Thebe, Io, Europa, Ganymede,
Callisto and Himalia. Our algorithm predicts two missing moons between
the two outermost detected moons Callisto and Himalia
(similar to Fig.~6 in Aschwanden and McFadden 2017).

In panel Saturn-10 (in Fig.~2c) we sample all Saturn moons with a
diameter of $d > 100$ km, namely Janus, Mimas, Enceladus, Thetis,
Dione, Rhea, Tian, Hyperion, Japetus, and Phoebe.
Our algorithm predicts one missing moon between
the two outermost detected moons Japetus and Phoebe
(similar to Fig.~7 in Aschwanden and McFadden 2017).

In panel Uranus-8 (in Fig.~2c) we sample all Urnaus moons with a
diameter of $d > 100$ km, namely Portia, Puck, Miranda, Ariel,
Umbriel, Titania, Oberon, and Scycorax.
Our algorithm predicts three missing moons between
the two outermost detected moons Oberon and Scycorax
(similar to Fig.~8 in Aschwanden and McFadden 2017).

In panel Neptune-6 (in Fig.~2c) we sample all Neptune moons with a
diameter of $d > 100$ km, namely Galatea, Despina, Larissa, Proteus,
Proteus, Triton, and Nereid.
Our algorithm predicts three missing moons between
the two outermost detected moons Triton and Nereid,
and one missing moon between Proteus and Triton 
(similar to Fig.~9 in Aschwanden and McFadden 2017).

Altogether, our harmonic resonance model with an upper limit of
$Q=3.0$ for the highest admissible harmonic number predicts a
total of 10 missing moons in our own planetary system. 
Note, that none of the predicted sequences exceeds a maximum
of 11 moons per planetary system. 

\bigskip
\acknowledgements
This research has made use of the Exoplanet Orbit Database and the
Exoplanet Data Explorer at {\sl exoplanets.org}.
The first author acknowledges the hospitality and partial support for 
two workshops on ``Self-Organized Criticality and Turbulence'' at the
{\sl International Space Science Institute (ISSI)} at Bern, Switzerland,
during October 15-19, 2012, and September 16-20, 2013, as well as
constructive and stimulating discussions (in alphabetical order)
with Sandra Chapman, Paul Charbonneau, Henrik Jeldtoft Jensen,
Maya Paczuski, Jens Juul Rasmussen, John Rundle, Loukas Vlahos,
and Nick Watkins.
This work was partially supported by NASA contract NNX11A099G
``Self-organized criticality in solar physics''.

\clearpage


\clearpage

\begin{deluxetable}{lr}
\tablecaption{Statistics of analyzed data set, using the Exoplanet
Orbit Database (EOD) at exoplanets.org, (status of 2017 May 10).}
\tablewidth{0pt}
\tablehead{
\colhead{Data set}&
\colhead{Number}}
\startdata
Number of detected exoplanets 			& 5454 \\
Number of exoplanet associated stars 		& 4522 \\
Number of stars with 1 exoplanet 		& 3860 \\
Number of stars with 2 exoplanets 		&  472 \\
Number of stars with 3 exoplanets 		&  128 \\
Number of stars with 4 exoplanets 		&   44 \\
Number of stars with 5 exoplanets 		&   14 \\
Number of stars with 6 exoplanets 		&    2 \\
Number of stars with 7 exoplanets 		&    2 \\
						&      \\
Number of planet-pair orbit ratios		&  932 \\
\enddata
\end{deluxetable}
\clearpage


\begin{figure}
\plotone{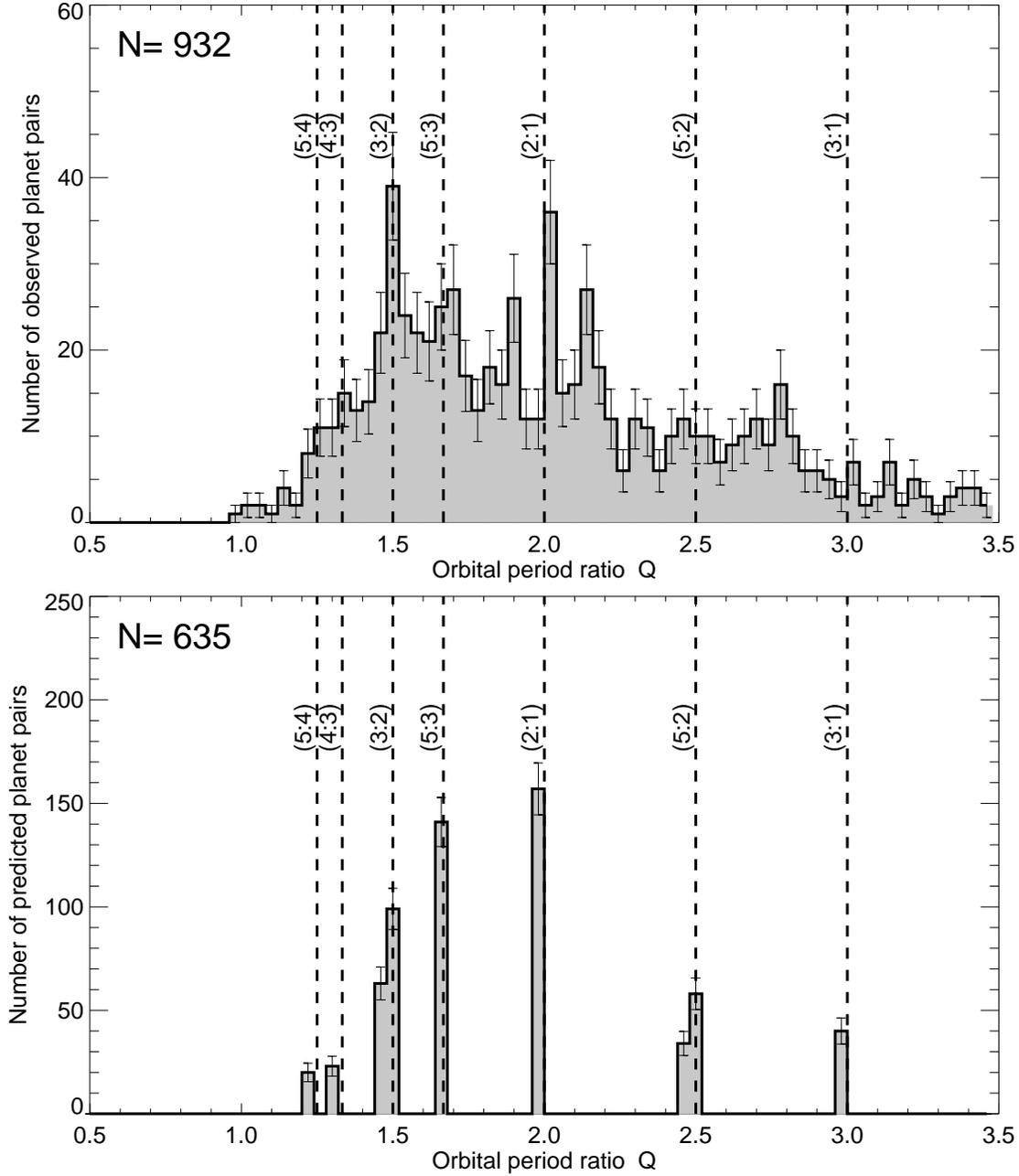}
\caption{{\sl Top:} The distribution of orbital period ratios $Q$ 
in all 932 cases of planet pairs measured in the exoplanet
database, using the version of 2017-May-10, which contains
confirmed planets as well as candidate planets from the Kepler
mission. The error bars conform to Poisson statistics. The 7 principal  
harmonic ratios are marked with vertical dashed lines.
{\sl Bottom:} The predicted quantized distribution of orbital 
period ratios $Q$ in complete (gap-free) sequences of exoplanet 
detections (with at least 3 detections), obtained after filling 
gaps with $Q > 3.0$ with a search strategy described in Section 2.4.}
\end{figure}

\begin{figure}
\plotone{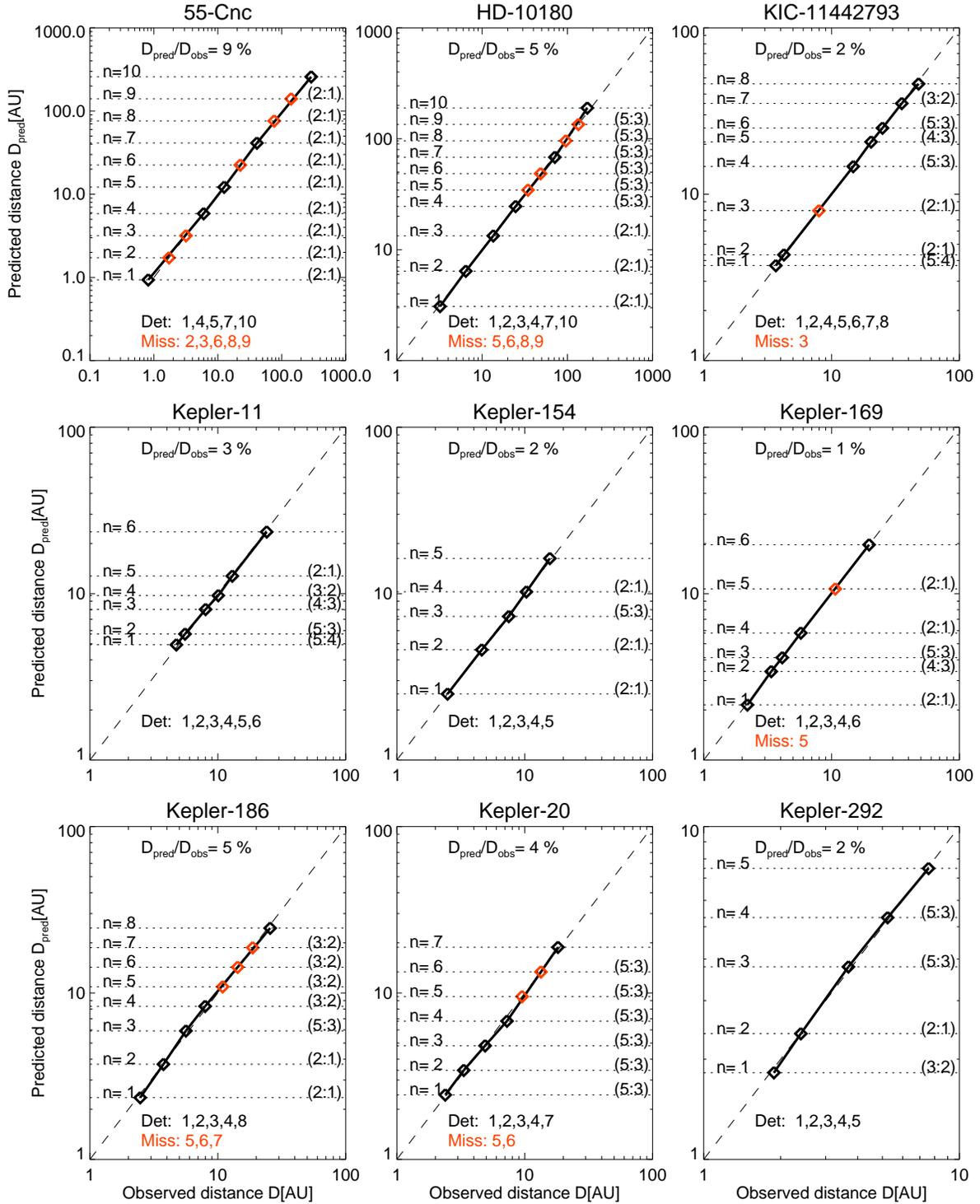}
\caption{The predicted distances $D^{pred}$ versus the observed
distances $D^{obs}$ of exoplanets from their central star in
large exoplanet systems (with 5, 6, or 7 members). The detected
exoplanets are marked with black diamonds, and the predicted
exoplanets with red diamonds. The numeration corresponds to the
complete systems that includes the predicted exoplanets. 
The average deviation between the prediced and observed distances,
$D_{pred}/D_{obs}$, are given in percentages, the best-fit harmonic
ratios are given on the right side, and the numeration of the
detected and missing planets (red) are indicated near the bottom
of each panel.}
\end{figure}

\begin{figure}
\plotone{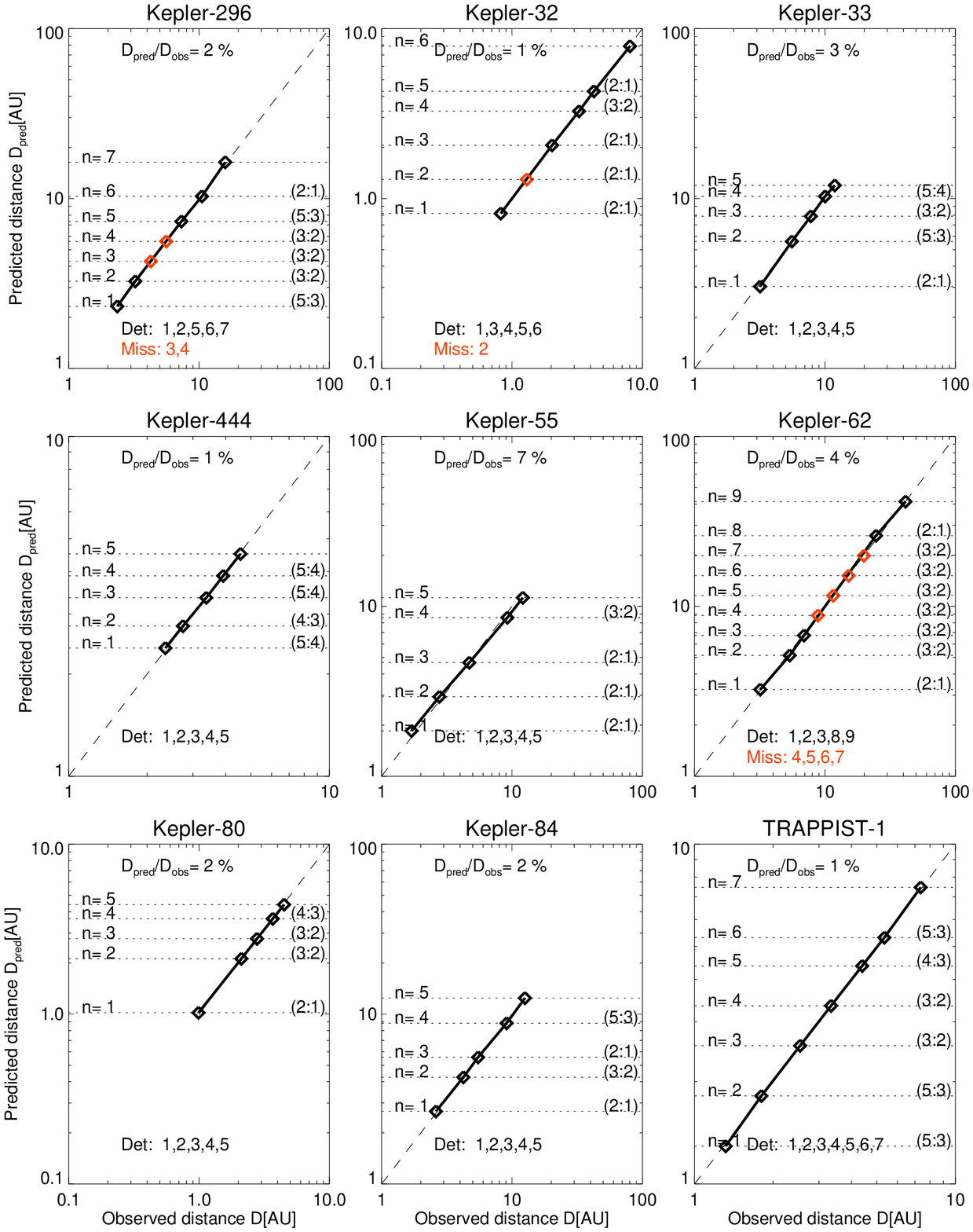}
\end{figure}

\begin{figure}
\plotone{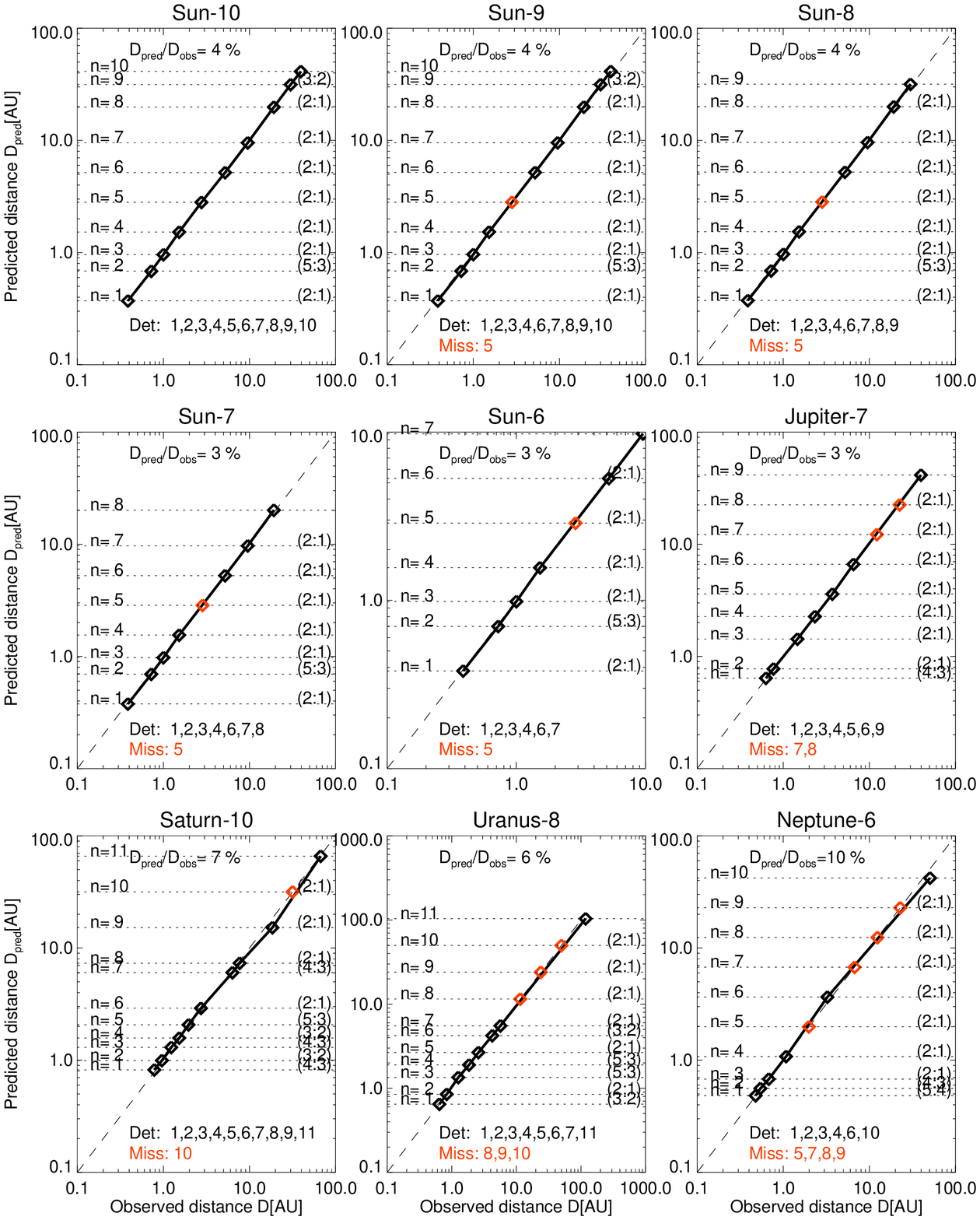}
\end{figure}

\begin{figure}
\plotone{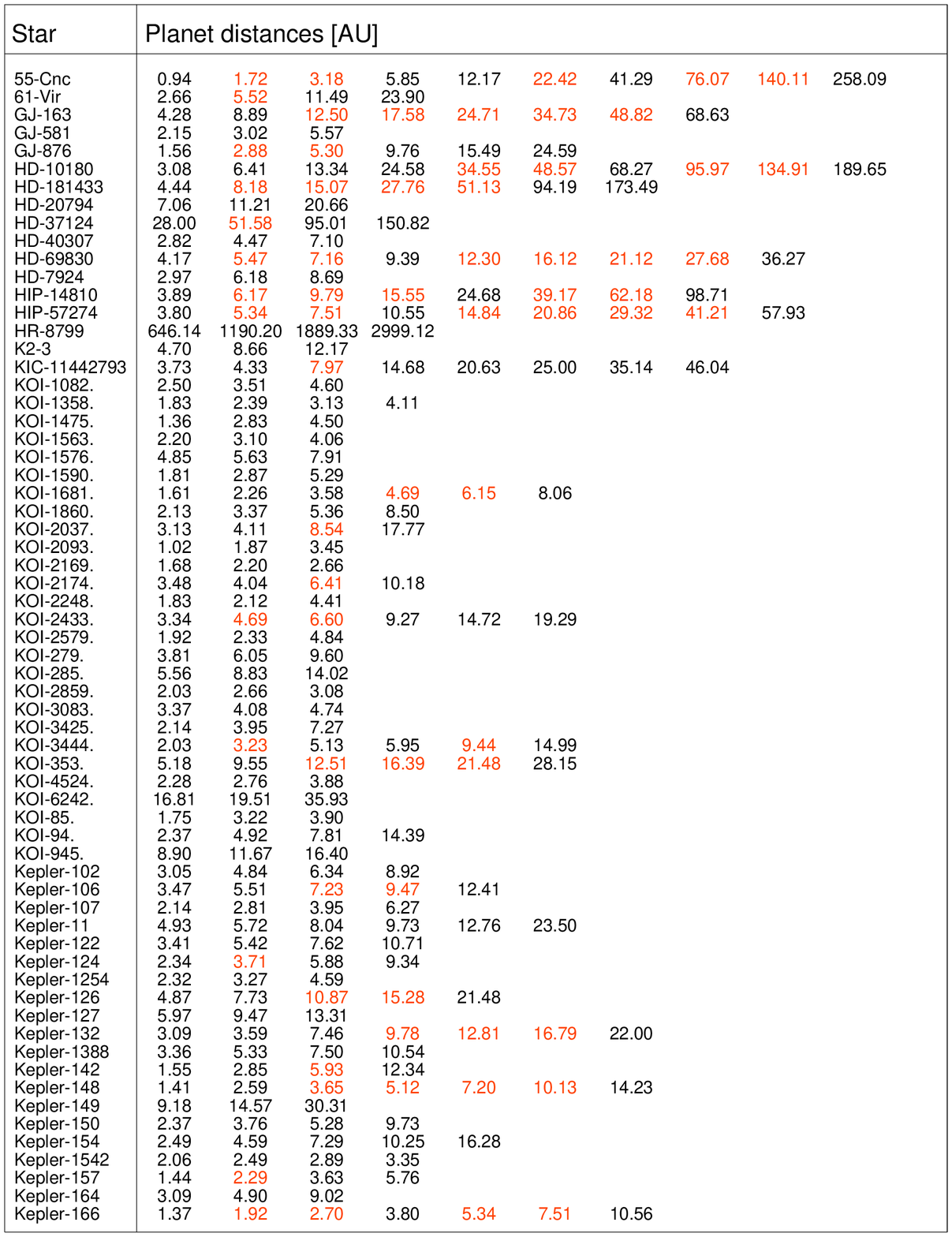}
\caption{Observed (black) and predicted (red) distances of exoplanets.}
\end{figure}

\begin{figure}
\plotone{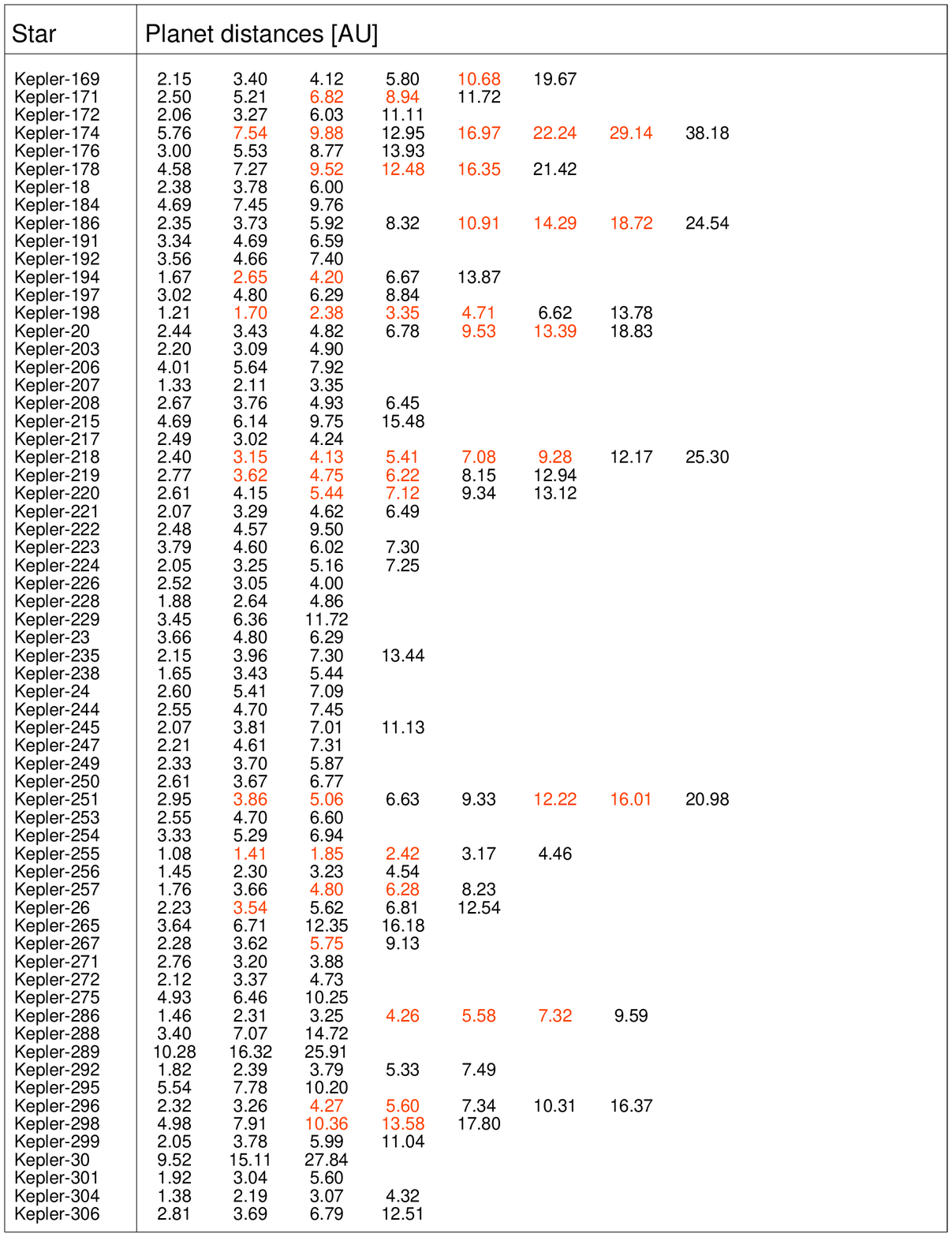}
\end{figure}

\begin{figure}
\plotone{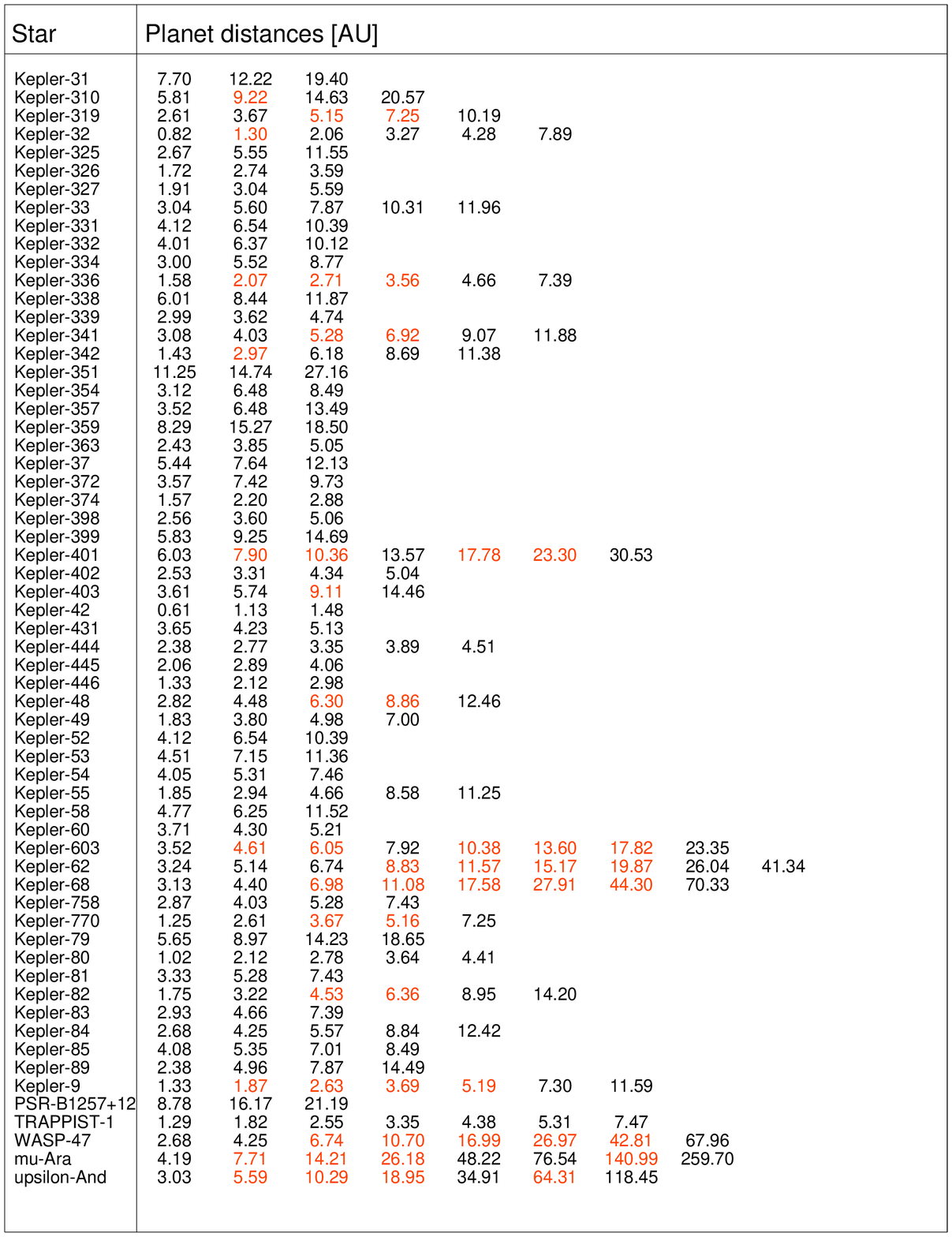}
\end{figure}


\begin{references} 

\def\ref#1{\par\noindent\hangindent1cm {#1}}

\ref{Aschwanden, M.J., Crosby, N., Dimitropoulou, M., Georgoulis, M.K., 
	Hergarten, S., McAteer, J., Milovanov, A., Mineshige, S., Morales, L., 
	Nishizuka, N., Pruessner, G., Sanchez, R., Sharma, S., Strugarek, A., 
	and Uritsky, V. 2016, Space Science Reviews 198, 47.}
\ref{Aschwanden, M.J. and McFadden, L.A. 2017,
 	``Harmonic Resonances of Planet and Moon Orbits - From the 
	Titius-Bode Law to Self-Organizing Systems'',
 	URL=http://www.lmsal.com/$\sim$aschwand/eprints/ 2017$\_$planets.pdf, 
 	arXiv:1701.08181 astro-ph.}
\ref{Bovaird, T. and Lineweaver, C.H., 2013, MNRAS 435, 1126.}
\ref{Bovaird, T., Lineweaver, C.H., and Jacobsen, S.K.
	2015, MNRAS 448, 3608.}
\ref{Brown, E.W.  and Shook, C.A. 1933, {\sl Planetary Theory},
	reprinted 1965. New York: Dover.}
\ref{Chandler, C.O., McDonald, I., and Kane, S.R. 2016, ApJ 151, 3.}
\ref{Cuntz, M. 2012, PASJ 64, 73.}
\ref{Ghilea, M.C. 2105, arXiv:1410.2478v3.}
\ref{Gillon, M., Triaud, A.H.M.J., Demory, B.O., et al.
         2017, Nature 542, 456 (23 Feb 2017).}
\ref{Han, E., Wang, S., Wright, J.T., Feng, Y.K., Zhao, M., Fakhouri, O.,
	Brown, J.I., and Hancock, C. 2014, Pubs. Astron Soc. Pacific,
	Vol. 126, Issue 943, pp. 827.}
\ref{Huang, C.X. and Bakos, G.A. 2014, MNRAS 442, 674.}
\ref{Lovis, C., Segransan, D., Mayor, M., Udry, S., Benz, W.,
        Bertaux, J.L., Bouchy, F., Correia, A.C.M., et al.
        2011, A\&A 528, A112.}
\ref{Peale, S.J. 1976, ARAA 14, 215.}
\ref{Pletser, V. and Basano, L. 2017, arXiv preprint.}
\ref{Poveda, A. and Lara, P. 2008, Revista Mexicana de Astronomia
        y Astrofisica 44, 243.}
\ref{Qian, S.B., Liu, L., Liao, W.P., Li, J., Zhu, L.Y., Dai, Z.B.,
        He J.J., Zhao, E.G., Zhang, J., and Li, K.
	2011, MNRAS 414, L16.}
\ref{Scholkmann, F. 2013, Progress in Physics Vol. 9 (4), 85.}
\ref{Scholkmann, F. 2017, Progress in Physics Vol. 13, 125.}
\end{references}
\end{document}